\documentclass[twocolumn,aps,tightenlines,floatfix,showpacs]{revtex4}
\usepackage{graphicx}
\begin{document}

\title{Configuration Interactions Constrained by Energy Density Functionals}

\author{B. Alex Brown$^{1}$, Angelo Signoracci$^{1}$ and Morten Hjorth-Jensen$^{2}$}

\affiliation{$^{1}$Department of Physics and Astronomy,
and National Superconducting
Cyclotron Laboratory,
Michigan State University,
East Lansing, Michigan 48824-1321, USA}

\affiliation{$^{2}$Department of Physics and Center for Mathematical Applications,
University of Oslo,
N-0316, Oslo, Norway}

\begin{abstract}
A new method for constructing a Hamiltonian for configuration interaction
calculations with constraints to energies of spherical configurations
obtained with energy-density-functional (EDF) methods is presented. This results
in a unified model that reproduced the EDF binding-energy in the limit of
single-Slater determinants, but can also be used for obtaining energy
spectra and correlation energies
with renormalized nucleon-nucleon interactions. The three-body
and/or density-dependent terms that are necessary for good nuclear
saturation properties are contained in the EDF. Applications to
binding energies and spectra of nuclei in the region above $^{208}$Pb are given.
\end{abstract}

\maketitle

In nuclear structure theory the two main computational methods for heavy nuclei
based upon the
nucleon fermionic degrees of freedom are the Hartree-Fock or energy-density-functional
(EDF) method and the configuration interaction (CI) method.
The EDF method is often limited to a configuration with a single Slater determinant.
The EDF Hamiltonian has parameters that are fitted to global properties of nuclei such
as binding-energies and rms charge radii \cite{skx}, \cite{rein}.

The CI method takes into
account many Slater determinants. CI often uses a Hamiltonian derived from
experimental single-particle energies and a microscopic
nucleon-nucleon interaction \cite{cov}. A given CI Hamiltonian
is applied to a limited mass region
that is related to the configurations of a few valence orbitals
outside of a closed shell and the associated renormalized
nucleon-nucleon interaction
that is specific to that mass region \cite{morten}, \cite{cov}.
Spectra and binding energies (relative to the closed core)
obtained from such calculations for two to four valence
particles are in good agreement with experiment \cite{morten}, \cite{cov}.
As many valence nucleons are added the agreement with experimental
spectra and binding energies deteriorates \cite{bab}.
An important part that is missing from these CI calculations
is the effective two-body
interaction that comes from the three-body interaction
of two valence nucleons interacting with one nucleon in the core \cite{zuk}.
To improve agreement with experimental spectra
one often adjusts some of the valence two-body matrix elements.
The most important part of this adjustment can be
traced to the monopole component of the two-body matrix elements that
controls how the effective single-particle energies evolve as
a function of proton and neutron number \cite{zuk}.

\begin{figure}
\scalebox{0.5}{\includegraphics{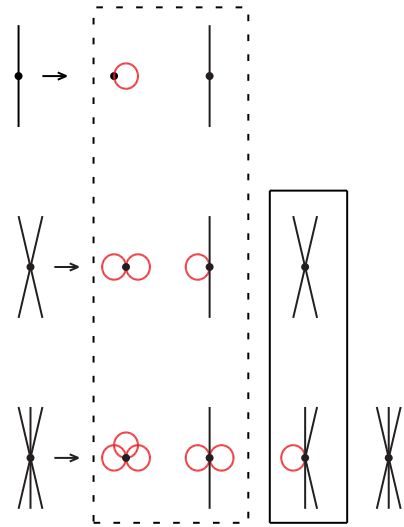}}
\caption{Schematic diagram for the terms in the Hamiltonian
obtained from Wick's theorem for a closed shell. The red lines
represent the summation over the orbitals in the closed shell.
The black lines represent the valence particles and/or holes.}
\label{(1)}
\end{figure}

Fig. 1 shows Wick's theorem applied to a closed shell
for the one-body kinetic energy, the two-body interaction
and the three-body interaction.
The part contained in the dashed box represents the closed-shell and
effective one-body
parts of the Hamiltonian that might be contained in an EDF approach.
Up to now this has been treated phenomenologically in the framework
of the Skyrme Hartree-Fock or relativistic Hartree method with
some parameters (typically 6-10) fitted to global experimental data.
There are efforts underway to relate the parameters of these
phenomenological approaches to the underlying two and three
body forces between nucleons, and also to extend the functional forms
to obtain improved agreement with experiment \cite{dft}. The part contained in
the solid-line box is the residual interaction used for CI calculations.
The remaining term is a valence three-body interaction.

In this paper we discuss a new method for obtaining
a valence Hamiltonian for valence nucleons outside
of a doubly-closed shell. The specific application
is made for $^{208}$Pb, but it could be
applied to any other doubly closed-shell system.
The single-particle
energy for orbital $  a  $ is defined as
$$
e_{a} = E(^{208}{\rm Pb}+a) - E(^{208}{\rm Pb}),       \eqno({1})
$$
where $  E(^{208}{\rm Pb})  $ is the energy of the
closed-shell configuration for $^{208}$Pb,
and $  E(^{208}{\rm Pb}+a)  $ is the energy of
the closed-shell configuration plus one nucleon constrained
to be in orbital $  a  $. Eq. 1 defines the one-body part of
the CI calculations. Often experimental data are used
for the energies in Eq. 1. In this paper we will use the
results of EDF calculations for these energies.
The practical use of Eq. 1 requires that two
states be connected by a spectroscopic factor
of near unity.

The two-body part of the CI Hamiltonian
is obtained with the usual renormalization
procedure \cite{morten}. For our examples, the active valence orbitals are
$  (0h_{9/2},1f_{7/2},0i_{13/2})  $ for protons and
$  (0i_{11/2},1g_{9/2},0j_{15/2})  $
for neutrons.
For the two-body valence interaction
we use the N$^{3}$LO nucleon-nucleon interaction \cite{n3lo}
renormalized to the nuclear medium with the V$_{lowk}$ method \cite{bogner}
with a cut-off of $  \Lambda =2.2  $ fm. Core-polarization corrections
are calculated in second-order
up to 6$\hbar\omega$ in the excitations energy.
We use harmonic-oscillator radial wavefunctions
with $\hbar\omega$ = 6.883 MeV.

The new aspect of our method is to take
the monopole part of the effective two-body interaction from
$$
\bar{V}_{ab} = E(^{208}{\rm Pb}+a+b) - E(^{208}{\rm Pb}) - e_{a} - e_{b},       \eqno({2})
$$
where $  E(^{208}{\rm Pb}+a+b)  $ is the spherical EDF energy of the
configuration for a closed shell plus two nucleons
constrained to be in orbitals $  a  $ and $  b  $.
This monopole interaction contains both the two and three
body terms shown by the solid-line box in Fig. 1
to the extent that
they are contained in the EDF phenomenology.
We modify the monopole
part of the microscopic
valence interaction to reproduce the results of Eq. 2. With this
modification, the CI calculations closely reproduce the EDF calculations
for single-Slater determinants, even when relatively many
valence nucleons are added. Thus, the CI calculations are constrained
to reproduce the trends of closed-shell energies and effective
single-particle energies obtained with the EDF. For our
model space orbitals, Eq. 2
involves about thirty configurations for two nucleons
(proton-proton, neutron-neutron and proton-neutron), but these
calculations in a spherical basis are computationally fast.

For this paper we
will use the EDF results based on the Skxm Skyrme interaction \cite{skx}.
An important property of Skxm is that the
experimental single-particle energies for the low-lying single-particle
states around $^{208}$Pb are reproduced with an rms deviation of about
300 keV. Skxm also has a reasonable value of the incompressibility (234 MeV).
We are not aware of any other Skyrme interaction that can do better
for the single-particle energies as defined by Eq. 1.
For the lowest
state for protons ($  0h_{9/2}  $ for $^{209}$Bi)
and neutrons ($  1g_{9/2}  $ for $^{209}$Pb), the difference
between experiment and theory can be reduced to on the
order of 20 keV
with only a small increase of  $\chi^{2}$=0.82 to $\chi^{2}$=0.89 for all
of the data considered in \cite{skx}. This is accomplished by using
a higher weight for these two data and requires a small adjustment
of the Skxm parameters.
Since the precise energies of these orbitals are important
for the results presented here, we use this new Skyrme interaction
called Skxmb. If we use Skxm or any other Skyrme interaction, our
conclusions are the same, but the deviation with experiment is
worse mainly because the single-particle energies are worse.
The binding energy of $^{208}$Pb with Skxmb is 1636.46 MeV
compared to the experimental value of 1636.45 MeV.

\begin{figure}
\scalebox{0.47}{\includegraphics{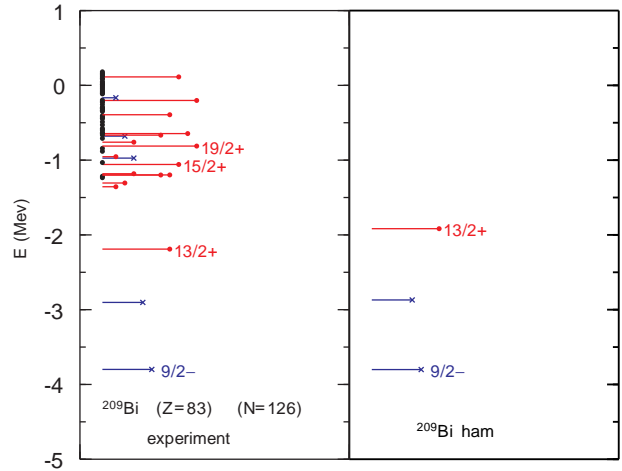}}
\caption{Comparison of experiment and theory (ham) for $^{209}$Bi.
The
energies are with respect to that of $^{208}$Pb.
The length of the lines indicate the spin with positive parity (red)
and negative parity (blue). Experimental levels that are unknown or
uncertain are shown by the black dots.}
\label{(2)}
\end{figure}
\begin{figure}
\scalebox{0.47}{\includegraphics{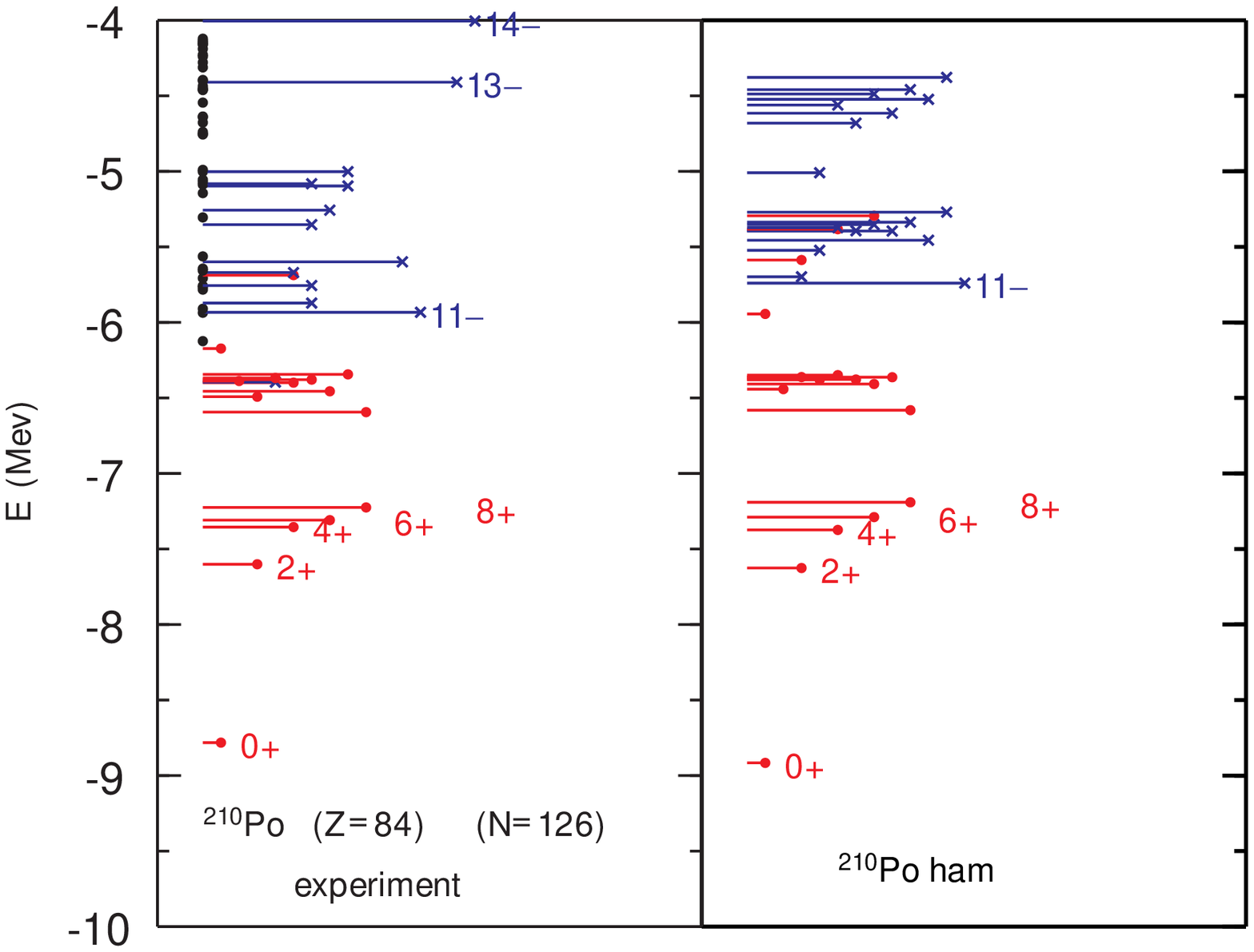}}
\caption{Comparison of experiment and theory (ham) for $^{210}$Po
(see caption to Fig. 2).}
\label{(3)}
\end{figure}
\begin{figure}
\scalebox{0.47}{\includegraphics{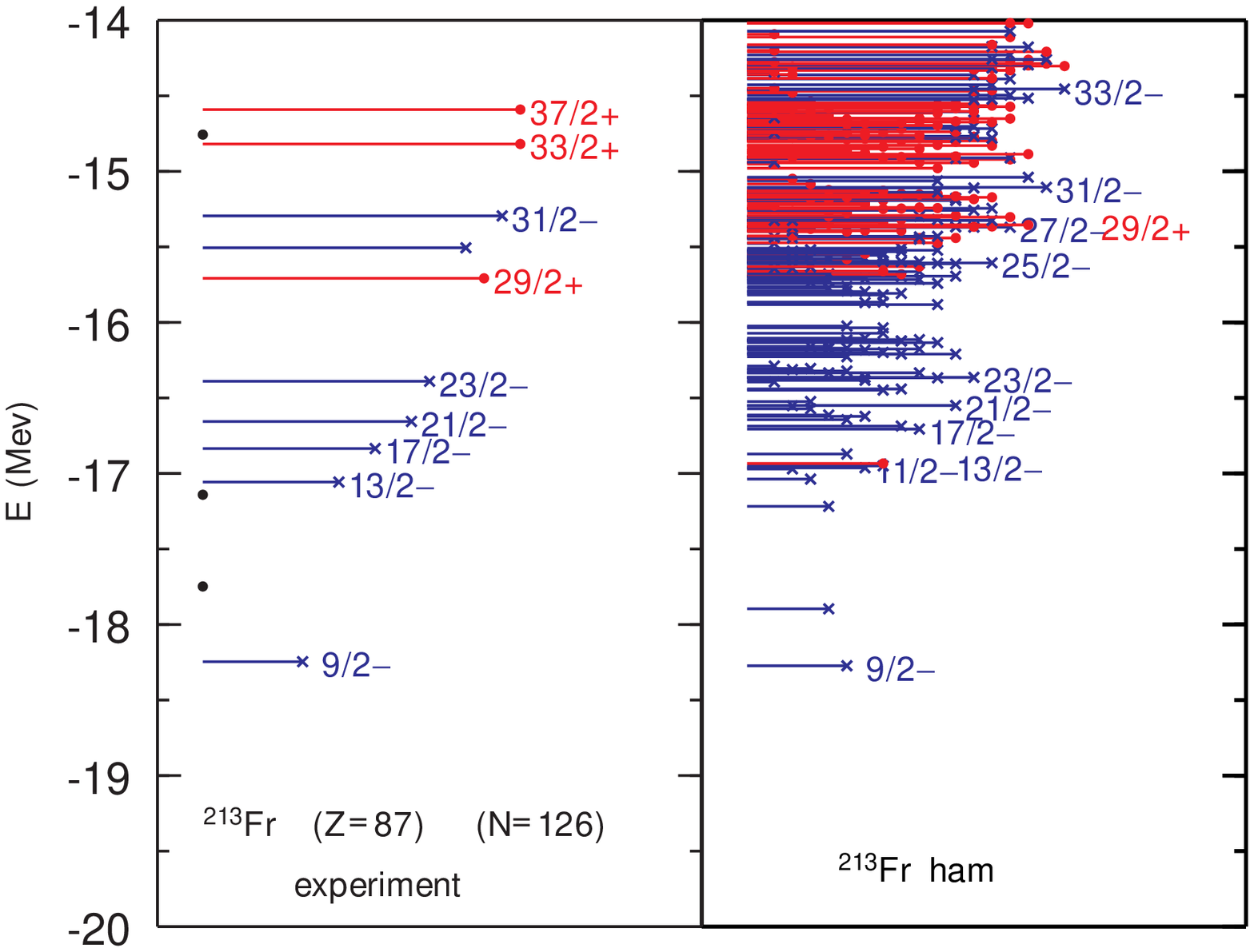}}
\caption{Comparison of experiment and theory (ham) for $^{213}$Fr
(see caption to Fig. 2).}
\label{(3)}
\end{figure}
\begin{figure}
\scalebox{0.47}{\includegraphics{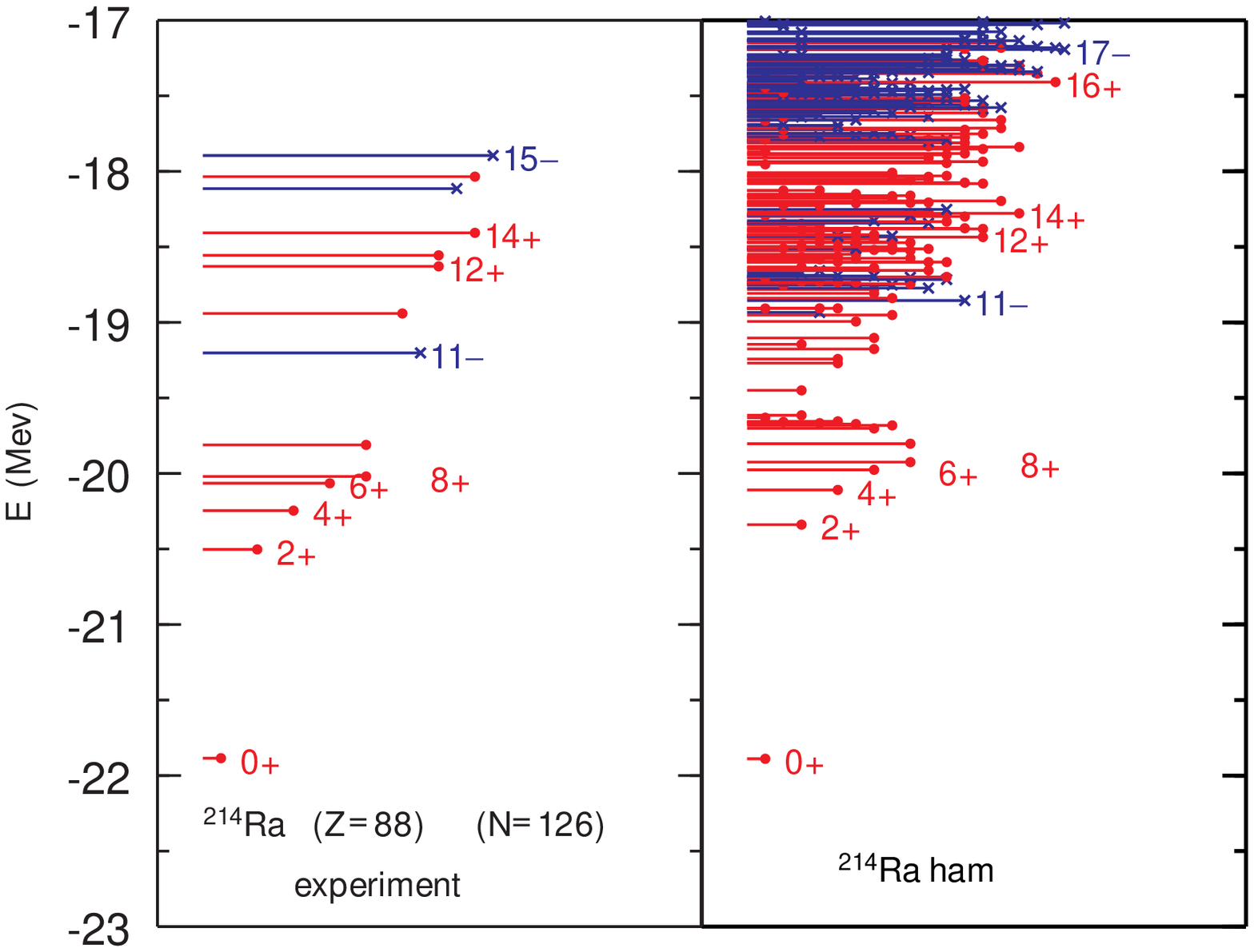}}
\caption{Comparison of experiment and theory (ham) for $^{214}$Ra
(see caption to Fig. 2).}
\label{(3)}
\end{figure}
\begin{figure}
\scalebox{0.47}{\includegraphics{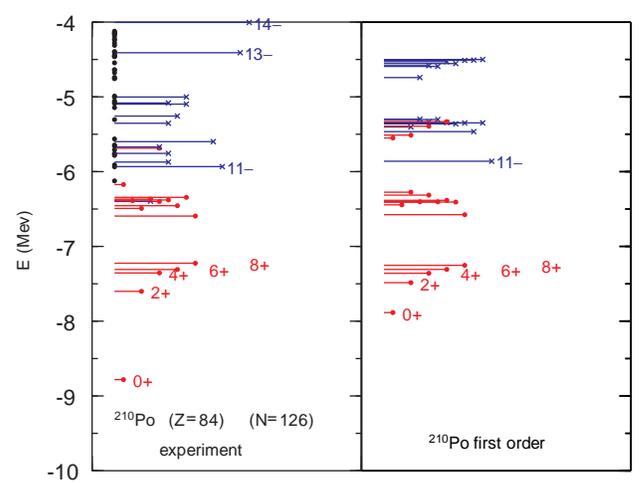}}
\caption{Comparison of experiment and theory (first order) for $^{210}$Po
(see caption to Fig. 2).}
\label{(3)}
\end{figure}

\begin{figure}
\scalebox{0.7}{\includegraphics{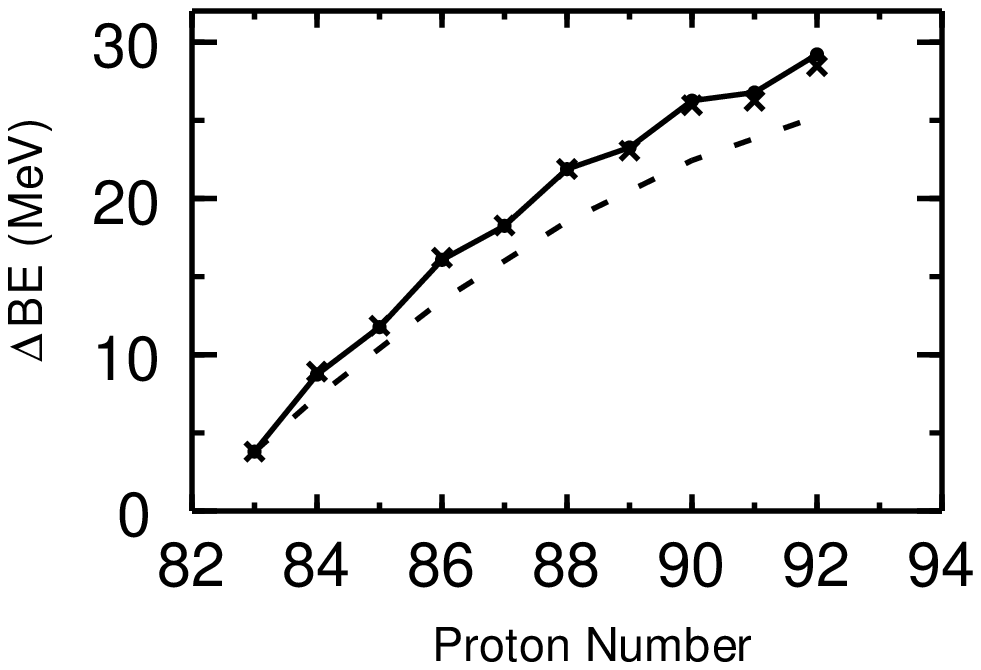}}
\caption{Binding energies relative to $^{208}$Pb. Experiment are
the points connected by a line. The results of the CI(N$^{3}$LO$+$Skxmb) are
shown by crosses. The results of the spherical EDF are shown by the
dashed line.}
\label{(3)}
\end{figure}

The results obtained from Skxmb for the
energies of single-particle states
$^{209}$Bi relative to the energy of $^{208}$Pb are
shown in Fig. 2. The energy of the lowest state, $  0h_{9/2}  $,
is reproduced due to the fit constraint. The next two states
(related to the $  1f_{7/2}  $ and $  0i_{13/2}  $ orbitals)
are also well reproduced.
One observes in experiment states related to core-excitation
of $^{208}$Pb starting about three MeV above the ground state.

For the lowest proton orbital with $  a=b=(  $0h$_{9/2}  )  $
the renomalized N$^{3}$LO monopole interaction is $  \bar{V}_{{\rm N^{3}LO}}  $=0.170 MeV
(it is repulsive due to the Coulomb interaction). The result obtained
from Eq. 2 with Skxmb is
$  \bar{V}_{{\rm EDF}}  $=0.288 MeV. The EDF-monopole
comes from both terms in the box in Fig. 1 and also
contains higher-order contributions implicit in the EDF
functional. Whereas, the N$^{3}$LO monopole only
contains the valence two-body interaction corrected to
second order. The difference is
$  \bar{V}_{{\rm EDF}} - \bar{V}_{{\rm N^{3}LO}}  $=0.118 MeV.
This correction is included in CI by modifying all of the
valence TBME $  <V>_{J}=\,<a b J\mid V\mid a b J>  $ for the $  0h_{9/2}  $ orbital by
$$
<\mid V\mid >_{J{\rm ,eff}} =
<\mid V\mid >_{J{\rm ,N^{3}LO}}
-\bar{V}_{{\rm N^{3}LO}}+\bar{V}_{{\rm EDF}}.       \eqno({3})
$$
Similar corrections are made for all other diagonal pairs of
orbital in the model space.

For the CI calculations we use the code NuShell \cite{nushell}.
The theory Hamiltonian (ham) consists of Skxmb for the
single-particle energies, and two-body matrix element
obtained from the renormalized N$^{3}$LO interaction
corrected to second-order, and then
finally the two-body monopoles corrected with Skxmb with Eq. 3.
The energies of $^{210}$Pb, $^{213}$Fr and $^{214}$Rn are shown in Fig. 3-5.
The agreement between experiment
and theory is good for the spectra and for the absolute energy relative
to $^{208}$Pb. For $^{210}$Po the agreement between experiment
and theory is very good for levels up to three MeV above the
ground state. Above three MeV the level density of
experiment and theory are similar, but one expects
additional levels in experiment coming from the core-excitation
of $^{208}$Pb. For  $^{213}$Fr and $^{214}$Ra
the theoretical level density is much
higher than experiment because the experimental
conditions select mainly the yrast levels.
For the low-lying levels in Figs. 3$-$5
the agreement between the absolute energies of experiment
and theory (relative to $^{208}$Pb) is usually within
100 keV, but there some exceptions with deviations
up to about 300 keV (e.g. the 11- in $^{214}$Ra).
These deviations
may be due to many factors
such as lack of third-order diagrams, the use of
the harmonic-oscillator basis for the renormalized
N$^{3}$LO matrix elements, non-monopole three-body
contributions, or inadequacies in the EDF Skxmb interaction.

When many nucleons are added, the monopole contribution goes as
$$
\Delta E = n(n-1)\bar{V}/2,       \eqno({4})
$$
where $  n  $ is the number of valence nucleons. Thus the
EDF monopole corrections become much more important as one adds
many valence nucleons.
When we constrain the CI to the single configuration
$  (0h_{9/2})^{10}  $ for the valence protons, the
CI calculation gives a binding energy increase of 25.05 MeV (relative
to $^{208}$Pb). The EDF calculation (with the same assumption for
the configuration) gives 25.24 MeV.
These are close to each other due to our EDF monopole correction to
the valence matrix elements. If the EDF monopole correction were
not included in CI the results would differ by (45)x(0.118) = 5.3 MeV.
The microscopic valance interaction on its own is too
strong and gives an ``over-saturation." The results for the
$  (1f_{7/2})^{8}  $ configuration are 13.27 MeV for CI and and 13.41 MeV
for EDF.
The difference between CI and EDF might be interpreted in
terms of an effective valence three-body monopole interaction
with strength $\Delta$E$_{3}$ = $  25.24 - 25.05 = 0.19  $ MeV for $  (0h_{9/2})^{10}  $
and $\Delta$E$_{3}$ = $  25.24 - 25.05 = 0.14  $ MeV $  (1f_{7/2})^{8}  $.
With $  \Delta E_{3}=n(n-1)(n-2)\bar{V}_{3}/6  $, $  \bar{V}_{3}  $ is on the
order of 1$-$2 keV.
$  \bar{V}_{3}  $ includes the three-body monopole interaction on the right-hand
side of Fig. 1, but it may also include other non-quadratic terms
that emerge from the EDF solutions.
For practical purposes $\Delta$E$_{3}$ is small
compared to other sources of error in the theory and it may be ignored.

The CI calculation were carried out up to $^{218}$U ($  Z=92  $)
where the M-scheme dimension is about 1.5 million.
The results for the ground state energies are compared to experiment
in Fig. 7.
The EDF calculation is based upon the spherical $  (0h_{9/2})^{n}  $ configuration
with $  n  $ = 1 to 10. The difference between EDF and CI can be regarded
as the correlation energy in the nuclear ground state, in this
case mainly due to the ``pairing" interaction. The correlation
results in wavefunctions that are highly mixed in the
valence proton basis. For example the ground state of
 $^{218}$U contains only 4.7\% of the $  (0h_{9/2})^{10}  $ component.
Up to $  Z=88  $ the difference between experiment and theory
for the binding energy relative to $^{208}$Pb
is on the order of 100 keV, and after this it gradually increases
to about 700 keV for $^{218}$U.

The pairing
interaction also appears in Fig. 3 for $^{210}$Po by the difference in energy
between the ground state and the J$^{ \pi }$=8$^{ + }$ state
which is dominated (99.86\%) by the $  (0h_{9/2})^{2}  $ configuration.
We show in Fig. 6 the
spectrum for $^{210}$Po obtained from the first-order N$^{3}$LO V$_{lowk}$
matrix elements. Comparison with Fig. 3 (which includes second
order) shows that two-thirds of the pairing comes from second-order
diagrams.
The tensor interaction is important for second-order pairing through
the bubble-diagram which links the valence protons with the core neutrons.

\begin{figure}
\scalebox{0.47}{\includegraphics{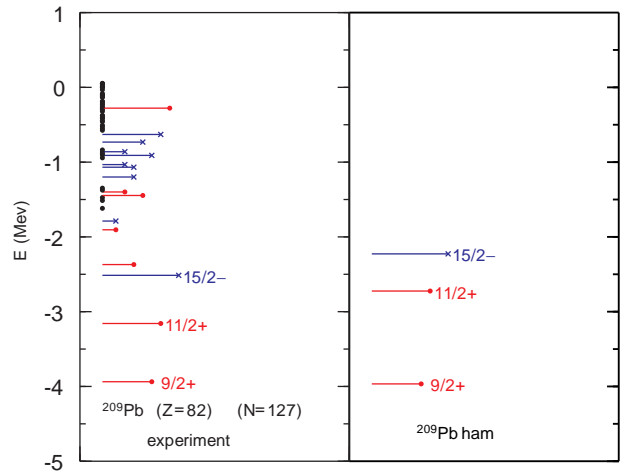}}
\caption{Comparison of experiment and theory (ham) for $^{209}$Pb
(see caption to Fig. 2).}
\label{(4)}
\end{figure}
\begin{figure}
\scalebox{0.47}{\includegraphics{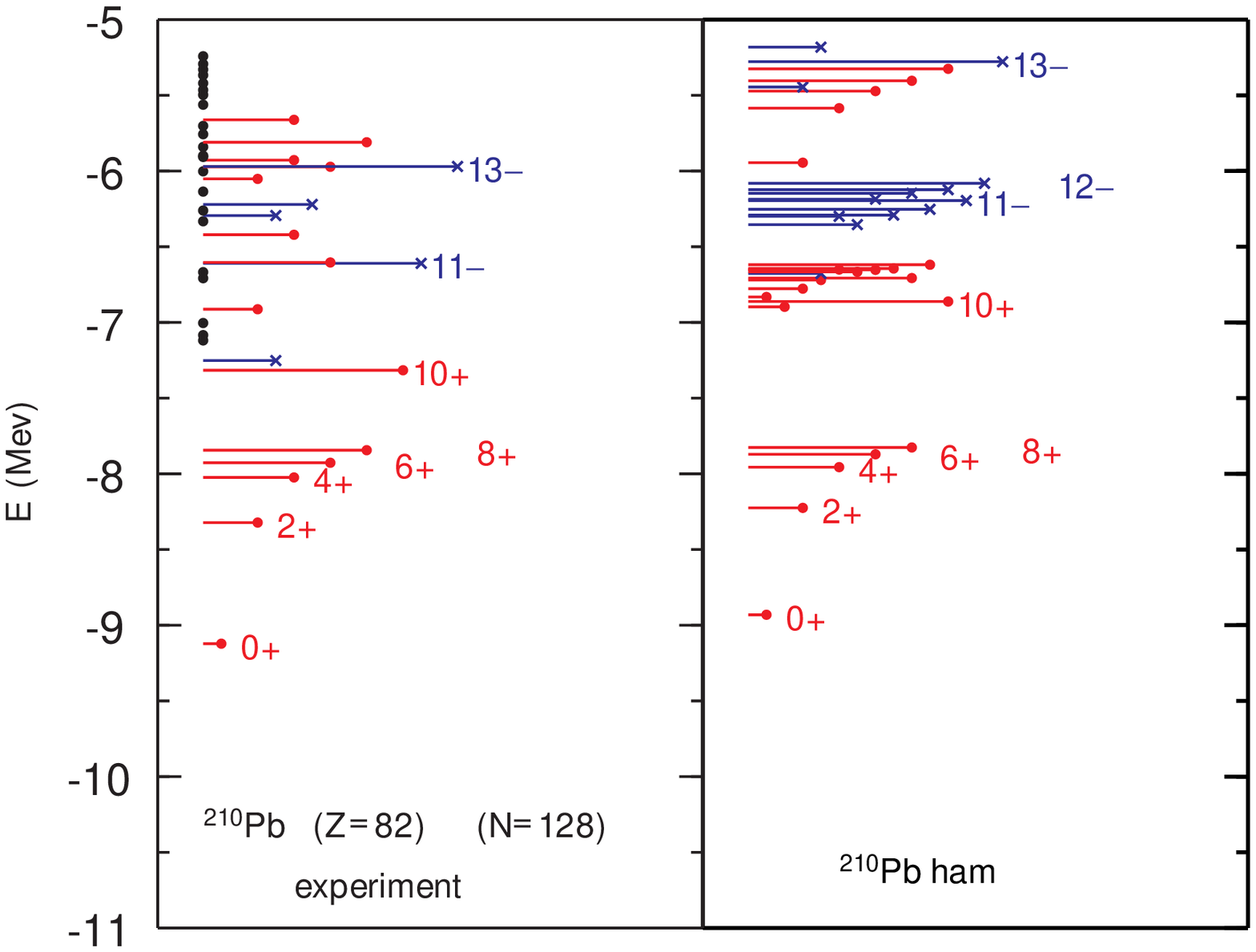}}
\caption{Comparison of experiment and theory (ham) for $^{210}$Pb
(see caption to Fig. 2).}
\label{(5)}
\end{figure}
\begin{figure}
\scalebox{0.47}{\includegraphics{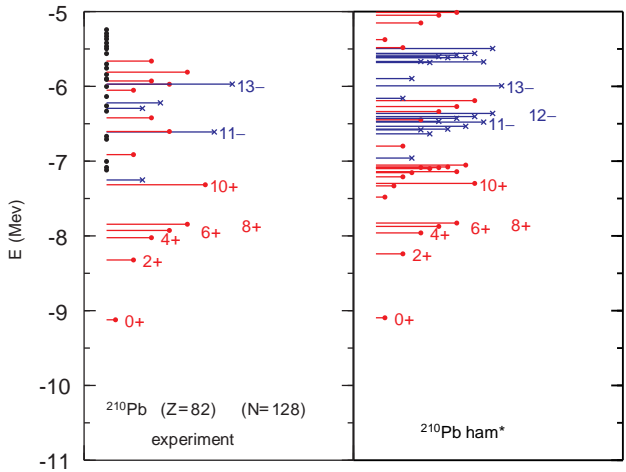}}
\caption{Comparison of experiment and theory (ham*) for $^{210}$Pb
(see caption to Fig. 2). Theory (ham*) is the same as theory (ham)
except that the single-particle energies for the neutron $  i_{11/2}  $
and $  j_{15/2}  $ orbitals are taked from the experimental values
in $^{209}$Pb.}
\label{(6)}
\end{figure}
\begin{figure}
\scalebox{0.47}{\includegraphics{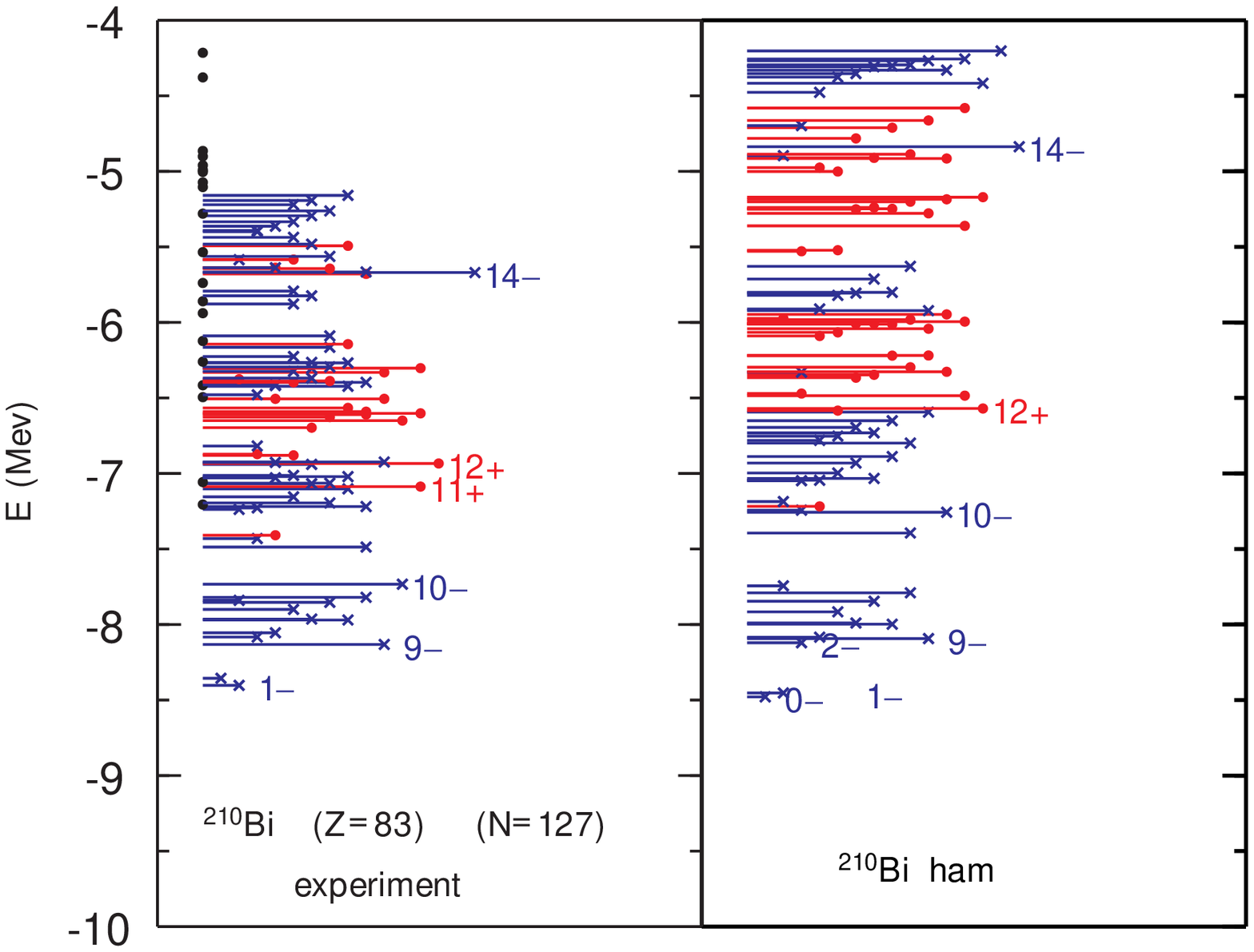}}
\caption{Comparison of experiment and theory (ham) for $^{210}$Bi
(see caption to Fig. 2).}
\label{(6)}
\end{figure}
\begin{figure}
\scalebox{0.47}{\includegraphics{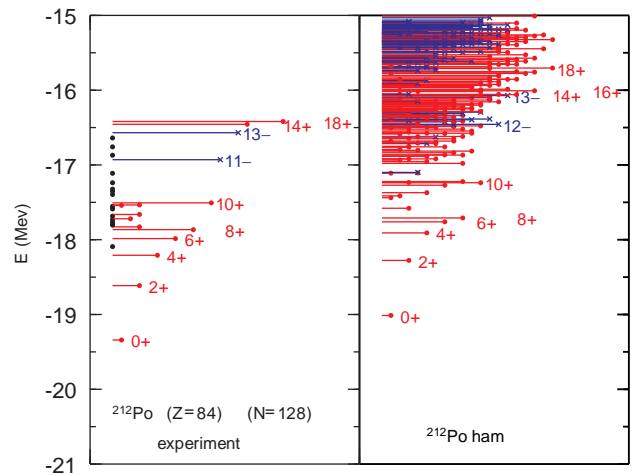}}
\caption{Comparison of experiment and theory (ham) for $^{212}$Po
(see caption to Fig. 2).}
\label{(6)}
\end{figure}

Results for neutrons for the spectra of $^{209}$Pb and $^{210}$Pb
are shown in Figs. 8 and 9, respectively. The single-particle energies
of the $  0i_{11/2}  $ and $  0j_{15/2}  $ orbitals in $^{209}$Pb
are 200-400 keV too
high with Skxmb. This is the reason why the theoretical
energies of the 10$^{ + }$, 11$^{-}$
and 13$^{-}$ states are too high in $^{210}$Pb. The results for $^{210}$Pb
are improved when the
energies of these two single-particle states are taken
from experiment
for $^{209}$Pb (left-hand side of Fig. 8) as shown in Fig. 10.
For our method to give the same results for CI and EDF
in the limit of spherical single Slater determinants,
one must take both the single-particle
energies and two-particle monopole energies from the
EDF calculation; one cannot arbitrarily change the single-particle
energies.   Thus it is important
to obtain EDF functionals that reproduce low-lying single-particle
energies near the doubly-magic nuclei.

Results
for $^{210}$Bi and $^{212}$Po are shown in Fig. 11 and 12.
Results for the low-lying proton-neutron spectrum of $^{210}$Bi
are comparable to those shown by \cite{cov}.
The theoretical
energies for the high-spin state would be in better agreement with
experiment if the experimental single-particle energies
from $^{209}$Pb are used for the neutrons.
But some specific disagreements remain, for example theoretical
the $  J^{\pi }=14^{-}  $ state in $^{210}$Bi remains about 500 keV too
high compared to experiment.

The monopole interactions for the $  1g_{9/2}  $ neutron orbital
are $  \bar{V}_{{\rm N^{3}LO}}  $= $-$0.076 MeV and $  \bar{V}_{{\rm EDF}}  $=0.017 MeV
giving a correction of 0.017 $-$ (0.076) = 0.093 MeV.
The monopole interactions between the $  0h_{9/2}  $ proton
orbital and the $  1g_{9/2}  $ neutron orbital
are $  \bar{V}_{{\rm N^{3}LO}}  $= $-$0.143 MeV and $  \bar{V}_{{\rm EDF}}  $= $-$0.205 MeV
giving a correction of $-$0.205 $-$ ($-$0.143) = $-$0.062 MeV.
Although the EDF monopole corrections are generally positive (leading
to less binding), some are negative, as in the last example.
This is a result of the microscopic dependence on the
specific orbitals being considered and their overlaps
with the central proton and neutron densities. The values depend
on the isoscalar and isovector properties
of the EDF functional that have parameters tuned to
reproduce global properties of binding energies.

In conclusion, we have provided a new method that is able to
constrain the monopole part of CI calculations to the EDF results in the
limit of single-Slater determinants. This constrained CI contains
all monopole interactions implicit in EDF including three-body,
density-dependent and rearrangement contributions.
In the limit of spherical single Slater determinants the CI
calculations with this method reproduces the EDF results except
for a very small three-body residual.
The results for the $  N=126  $ isotones show that
this change in the monopole interaction is crucial
for obtaining the correct absolute binding energies.
Second-order corrections are important for the
pairing interaction.
As illustrated in the case of $^{209}$Pb and $^{210}$Pb,
the accuracy of this method based on EDF results
for the monopole energies plus N$^{3}$LO for the
renormalized residual interaction
is limited by the
accuracy of the EDF methods to reproduce the binding
energies for states one nucleon removed from a closed shell
(Eq. 1).
In our examples for $^{208}$Pb the Skyrme parameters were
optimized for the precise ground-state energies of
of $^{209}$Bi and $^{209}$Pb leaving the rms
deviation for all other nuclei about the same
as shown in \cite{skx}. This method can be applied
to any other doubly-closed shell system, but its
accuracy will be limited by the accuracy of the EDF
results for single-particle energies.
Similar local optimizations may
be possible for other mass regions.
In the coming
years we may expect improvements in EDF theory and phenomenology
towards a improved universal functional.
For cases where the basis dimensions are too large
for exact CI methods,
it would be interested to apply our Hamiltonian
to approximate methods within this model space
for valence nucleons outside of $^{208}$Pb.

\vspace{0.2cm}
\noindent 
  {\bf Acknowledgments} This work is partly supported by NSF Grant
PHY-0758099 and the DOE UNEDF-SciDAC grant DE-FC02-09ER41585.

\end{document}